\documentclass[intlimits,twoside,a4paper]{article}

\usepackage[cp1251]{inputenc}
\usepackage[eqsecnum]{cmpj3}

\usepackage{bm}

\issue{2020}{23}{3}{33703}
\doinumber{10.5488/CMP.23.33703}
\title{Hybrid functional analysis of porous coordination polymers Cu[Cu(pdt)$_{2}$]  and Cu[Ni(pdt)$_{2}$]}
\author[S.V. Syrotyuk,  Yu.V. Klysko]{S.V. Syrotyuk,
        Yu.V. Klysko}
\address{
Department of Semiconductor Electronics, Lviv Polytechnic National University, \\1 S. Bandera St., 79013, Lviv, Ukraine
}
\date{Received November 20, 2019, in final form April 15, 2020}
\begin{document}

\maketitle

\begin{abstract}
Ab initio investigation of the two porous coordination polymers Cu[Cu(pdt)$ _{2} $] and Cu[Ni(pdt)$ _{2} $] has been performed. The dispersion laws and partial density of states was obtained with the PBE0 hybrid functional. The results found here show that the materials under consideration are degenerate $p$-type semiconductors. Here, the effect of partial self-interaction removing of the strongly correlated 3$d$ electrons of Cu and Ni was examined. In case of  Cu-containing materials, the obtained results confirm that the 3$d$ electrons of Cu reveal strong correlations, and, therefore, their electronic properties could be evaluated by means of a hybrid functional of the exchange-correlation energy.
We also obtained quasiparticle properties within the Green's function (G0W0) and Bethe-Salpeter approaches. The last one was used in order to examine excitonic properties in the degenerate semiconductors. The imaginary part of the dielectric function was obtained within random-phase approximation as well as the Bethe-Salpeter approach.

\end{abstract}

\section{Introduction}

The recently synthesized two metal-organic frameworks Cu[Cu(pdt)$ _{2} $] and Cu[Ni(pdt)$ _{2} $] are  characterized by tetragonal crystal structure, 1$d$ pore and relatively high conductivity for the group and high redox activity \cite{doi:10.1021/ic802117q, doi:10.1021/cm101238m}. Those properties make them attractive for the use in electronics as fuel or storage cells, electrodes or detectors.
Considering two works \cite{doi:10.1021/ic802117q, doi:10.1021/cm101238m}, where those materials were obtained for the first time, we can resume the following properties: both materials are $p$-type semiconductors, the substitution of Cu on Ni leads to an increase of optical gap from around 1.5 eV in Cu[Cu(pdt)$ _{2} $] to 2 eV in Cu[Ni(pdt)$ _{2} $].  There is also a possibility to enhance the conductivity of Cu[Ni(pdt)2] through partial oxidation with I$_{2} $ as oxidant. A broad peak at approximately 0.7 eV is observed in the absorption spectrum of Cu[Cu(pdt)$ _{2} $].

It is interesting that in case of Cu[Cu(pdt)$ _{2} $], the conductivity is higher for amorphous phase than for the crystalline \cite{C6CC09310H}. This is caused by changes of functional groups in the structure and by the generation of new Cu-S bonds.

Recent investigations concern the absorption and selectivity properties to gases and hydrocarbons.
Excellent water stability over a broad pH range as well as outstanding selectivity for C$ _{2} $H$ _{2} $/CO$ _{2} $ and C$ _{2} $H$ _{2} $/CH$ _{4} $ were defined in Cu[Ni(pdt)$ _{2} $] \cite{doi:10.1002/anie.201806732}.
It  was also shown that Cu[Ni(pdt)$ _{2} $] is an excellent adsorbent to separate propyne and propadiene from propylene \cite{doi:10.1002/anie.201904312}.
The prospect of using Cu[Ni(pdt)$ _{2} $] was proved by the investigation of the impact of gaseous hydrocarbons where an  impressive adsorption of ethane, ethylene, acetylene, propane, propylene, and cis-2-butene  was revealed as well as a strong impact on conductivity was observed \cite{doi:10.1021/jacs.9b00654}.
Let us now turn our attention to the choice of the method of study, keeping in mind the fact that there are two electron subsystems in the materials under consideration, namely, the $s(p)$ weakly and 3$d$ strongly correlated electrons, respectively.

Obviously, the local LDA and quasilocal GGA approaches are not effective for describing materials containing strongly correlated electrons. The GGA+U approximation is the first step for removing the self-interaction error (SIE) of the 3$d$ electrons, resulting in better locating the 3$d$ energy levels and  significantly improving the band gap. The shortcomings of this approach are as follows. First, the Hubbard parameter $U$ is system dependent, and there are no reliable methods for its determining. Second, each value of $U$ must be suplemented with three values of the screened Slater integrals, namely $ F^{0} $, $ F^{2} $ and $ F^{4} $, that is, for six nickel atoms in the material, you need to have 6$U$ parameters and 18 Slater integrals as input. The presence of 24 parameters leads to enormous difficulties in the numerical implementation of the electronic energy spectrum problem. Third, the GGA+U approach is considered to be a mean-field method based on one-site screened Coulomb energy, which is not updated in the iteration loops.

By contrast, the PBE0 method is based on the wave functions of 3$d$ electrons, which are updated in the process of self-consistent solution of the electron eigenvalue problem. It depends only on a single parameter, named the mixing factor, which will be described herein below. Unlike GGA+U, the PBE0 approach operates with a wave function and the potential of 3$d$ electrons that both change from iteration to iteration. 
The aim of our work is to evaluate the electronic structure of Cu[Cu(pdt)$ _{2} $] and Cu[Ni(pdt)$ _{2} $] by performing ab initio calculations with the hybrid functional in order to estimate the impact of  removing the SIE from the exchange part of the exchange-correlation energy of strongly correlated 3$d$ electrons.

\section{Methods}

All calculations were performed using ABINIT code \cite{abinit}, based on the projector augmented waves (PAW) \cite{PhysRevB.50.17953}. 
In the PAW formalism, every atom is characterized by an atom-centered augmentation sphere with a radius $ r_c $. Inside the sphere, a true all-electron wavefunction $ {\psi}_{n} $ is obtained from auxiliary nodeless smooth functionn $ \tilde{\psi}_{n} $:
\begin{equation}
| \psi_{n} \rangle = \hat{T} | \tilde{\psi}_{n} \rangle,
\label{eq:psi_psi_aux}
\end{equation}
where $ \hat{T} $ is a transformation operator, index $ n $ includes the wavevector \textbf{k} from the first Brillouin zone, the band and spin indexes. The Kohn-Sham equation, after taking into account equation (\ref{eq:psi_psi_aux}), is represented as follows:
\begin{equation}
\hat{T^{\dagger}} \hat{H} \hat{T} | \tilde{\psi}_{n} \rangle = \epsilon_{n} \hat{T^{\dagger}}\hat{T} | \tilde{\psi}_{n} \rangle.
\label{eq:KS_PAW}
\end{equation}

In the ABINIT code, the local exact exchange is implemented on the PAW basis. This is realized by mixing the exact Hartree-Fock exchange with the exchange-correlation functional GGA-PBE \cite{PBE} inside the atomic sphere determined by $ r_c $. This mixing results in the exchange-correlation hybrid functional PBE0, which is defined by the following equation \cite{doi:10.1063/1.478401, PhysRevB.74.155108}:
\begin{equation}
E_{xc}^{\text{PBE}0}[\rho] = E_{xc}^{\text{PBE}}[\rho] + \alpha(E_{x}^{\text{HF}}[\Psi_{3{d}}] - E_{x}^{\text{PBE}}[\rho_{3{d}}]),  
\label{eq:PBE0}
\end{equation}
where $ E_{xc}^{\text{PBE}}[\rho] $ is the GGA-PBE functional, $ E_{x}^{\text{HF}}[\Psi_{3{d}}] $ represents the exchange Hartree-Fock energy, and $ \Psi_{3{d}} $ and $ \rho_{3{d}} $ denote the wave function and the electron density of the strongly correlated   3$d$ electrons, respectively. In equation (\ref{eq:PBE0}), the coefficient $ \alpha $  determines a portion of the exact Hartree-Fock exchange in the exchange-correlation functional.

The GW calculations were performed  using the perturbative approach (one shot GW, i.e., G0W0). The G0W0 approach was used to define the so-called scissor energy operator  $ \Delta E_{g} $. 
The value $ \Delta E_{g} $ is determined as a difference between the G0W0 and GGA energy band gaps. It is used in the Bethe-Salpeter equation (BSE) for shifting the conduction band state energies.
Unfortunately, ABINIT does not support G0W0 calculations with the hybrid functionals in the PAW basis. We assume that $ \Delta E_{g} $ is the same for all ground state properties obtained with hybrid PBE0 and non-hybrid PBE functionals. Thus, the G0W0 approach was realized only on the eigenfunctions and eigenvalues obtained with the GGA-PBE functional. The BSE calculations were performed by direct diagonalization of the excitonic Hamiltonian only for a resonant block (Tamm-Dancoff approximation).
 
We used the experimental crystal structure obtained by the X-ray diffraction measurements \cite{doi:10.1021/ic802117q, doi:10.1021/cm101238m}. Elementary cell contains 44 atoms and is described by space group $ P4_{2/\text{mmc}} $ [131], and Bravais lattice is primitive tetragonal.

All calculations were performed with the following parameters: the integration in the momentum space was done with 18 $k$-points in the irreducible part of the Brillouin zone. The number of plane waves needed for the wave function expansion was defined by cut-off energy of 35 Ha. The electron density and crystal potential require a denser grid, which was generated by means of the cut-off energy of 150~Ha. The cut-off energy for a dielectric matrix was equal to 3.0~Ha. The limiting value of energy, used for evaluation of the exchange part of self-energy operator $\Sigma$, was equal to 60 Ha. All the parameters were chosen after many numerical experimentations in order to obtain the self-consistent electronic energy band structure and dielectric function of the materials under consideration.

\section{Results}
\subsection{Ground-state properties of Cu[Cu(pdt)$ _{2} $]}

\begin{figure}[!t]
\includegraphics[width=12cm]{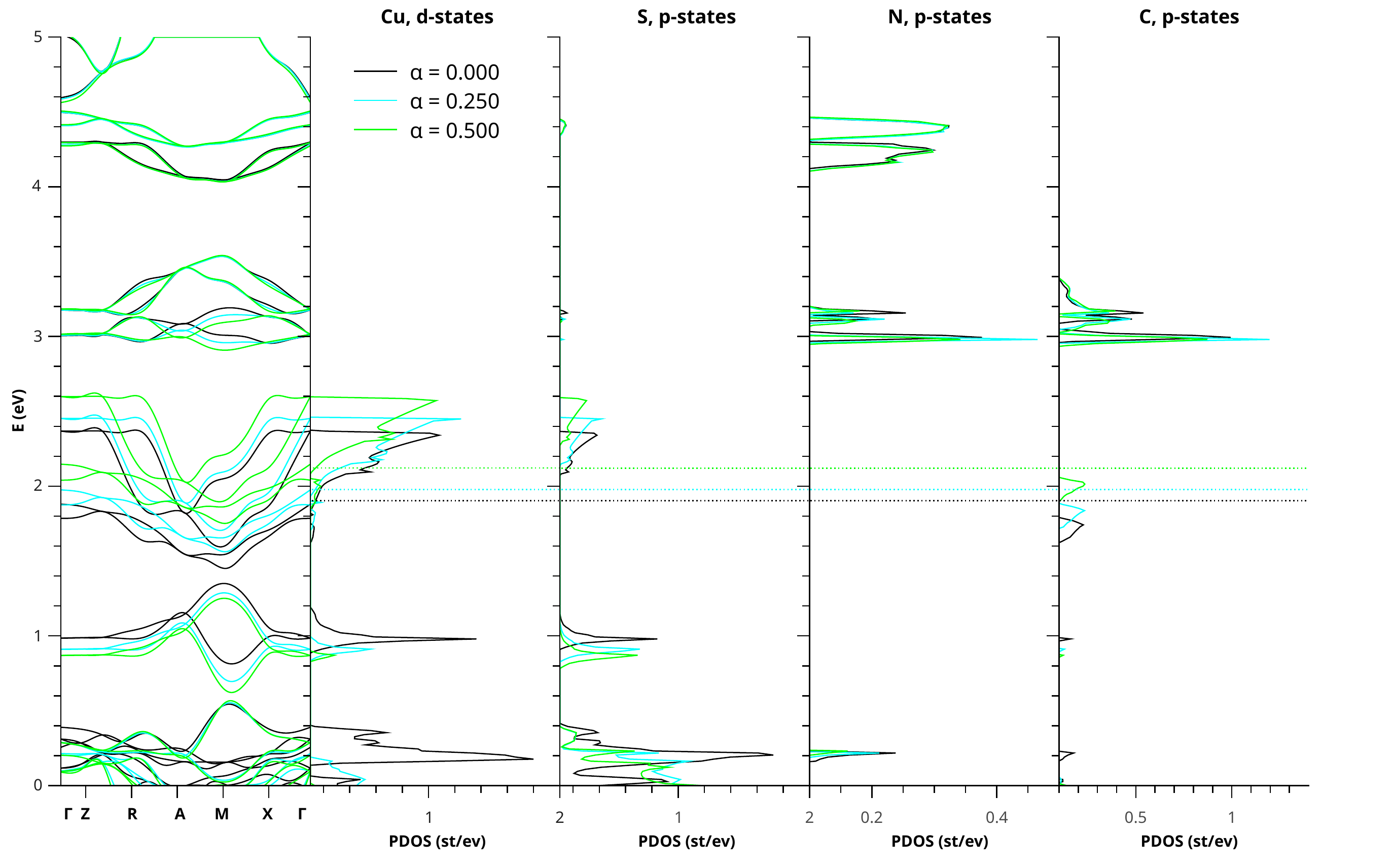}
\centering
\caption{(Color online) Electronic properties of Cu[Cu(pdt)$ _{2} $] including dispersion law and partial density of states (PDOS) obtained with several values of the parameter $ \alpha $. Dotted lines represent Fermi level.}
\label{fig:cu_cu_gs} 
\end{figure}

As can be seen from the ground-state calculations, the Fermi level lies at about 0.4 eV below the top of the valence band, populated by the Cu 3$d$ and S 3$p$ states, respectively (figure~\ref{fig:cu_cu_gs}). Therefore, we assume that the material Cu[Cu(pdt)$ _{2} $] is a $p$-type semiconductor. This hypothesis is confirmed by experimental data on its electrical conductivity, which is characteristic of semiconductors. The increase of $ \alpha $ results only in the energy shift of the valence bands without changes in the form of the energy dispersion curves. The dispersion curves in the conduction band do not reveal noticeable changes due to the absence of Cu 3$d$ states in this energy region. 

\begin{figure}
\includegraphics[width=10cm]{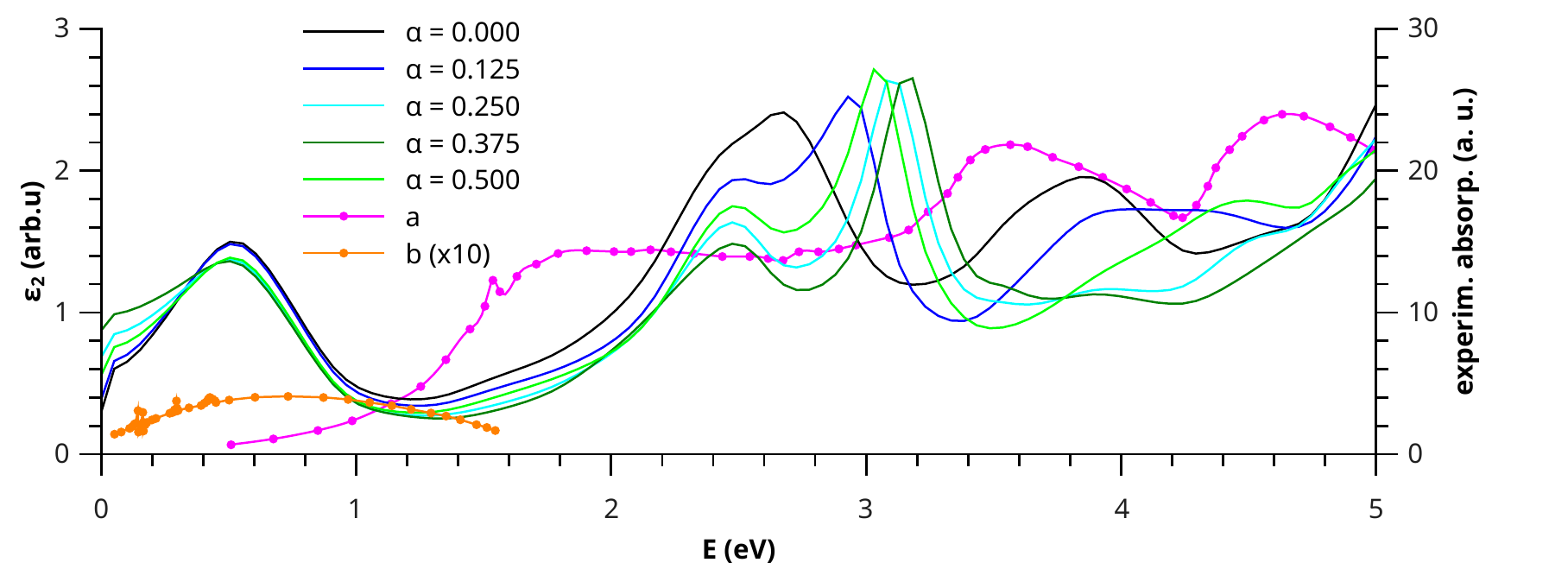}
\centering
\caption{(Color online) Imaginary part of the dielectric function $ \varepsilon_2^{\text{RPA}} $ for Cu[Cu(pdt)$ _{2} $] obtained within the RPA on base of the hybrid PBE0 functional, with several values of parameter $ \alpha $, in comparison with the experimental absorption spectra,
	a -- \cite{doi:10.1021/ic802117q}
	b -- \cite{doi:10.1021/cm101238m}.
}
\label{fig:cu_cu_opt} 
\end{figure}

An imaginary part of the RPA dielectric function $ \epsilon_2^{\text{RPA}} $ has a broad peak located at 0.5 eV (figure~\ref{fig:cu_cu_opt}). Its shape and energy is the same for all values of parameter $ \alpha $ used here.  This peak is supposed to be assigned to transitions between 3$d$ states of Cu and 3$p$ states of S which are located near the Fermi level (figure~\ref{fig:cu_cu_gs}). The energy dispersion curves show a very weak dependence on the mixing coefficient $ \alpha $. They are shifted in energy with an increase of a parameter $ \alpha $. The first absorption maximum agrees with the experimental one, observed at 0.7 eV. The disagreement between $ \varepsilon_2^{\text{RPA}} $ and the measured data is observed in the visible part of an absorption spectrum. A strong absorption peak is observed at photon energy of about 1.5 eV, when $ \varepsilon_2^{\text{RPA}} $ rises at the energies above 2 eV. 

\subsection{Ground-state properties of Cu[Ni(pdt)$ _{2} $]}

\begin{figure}[!t]
\includegraphics[width=\linewidth]{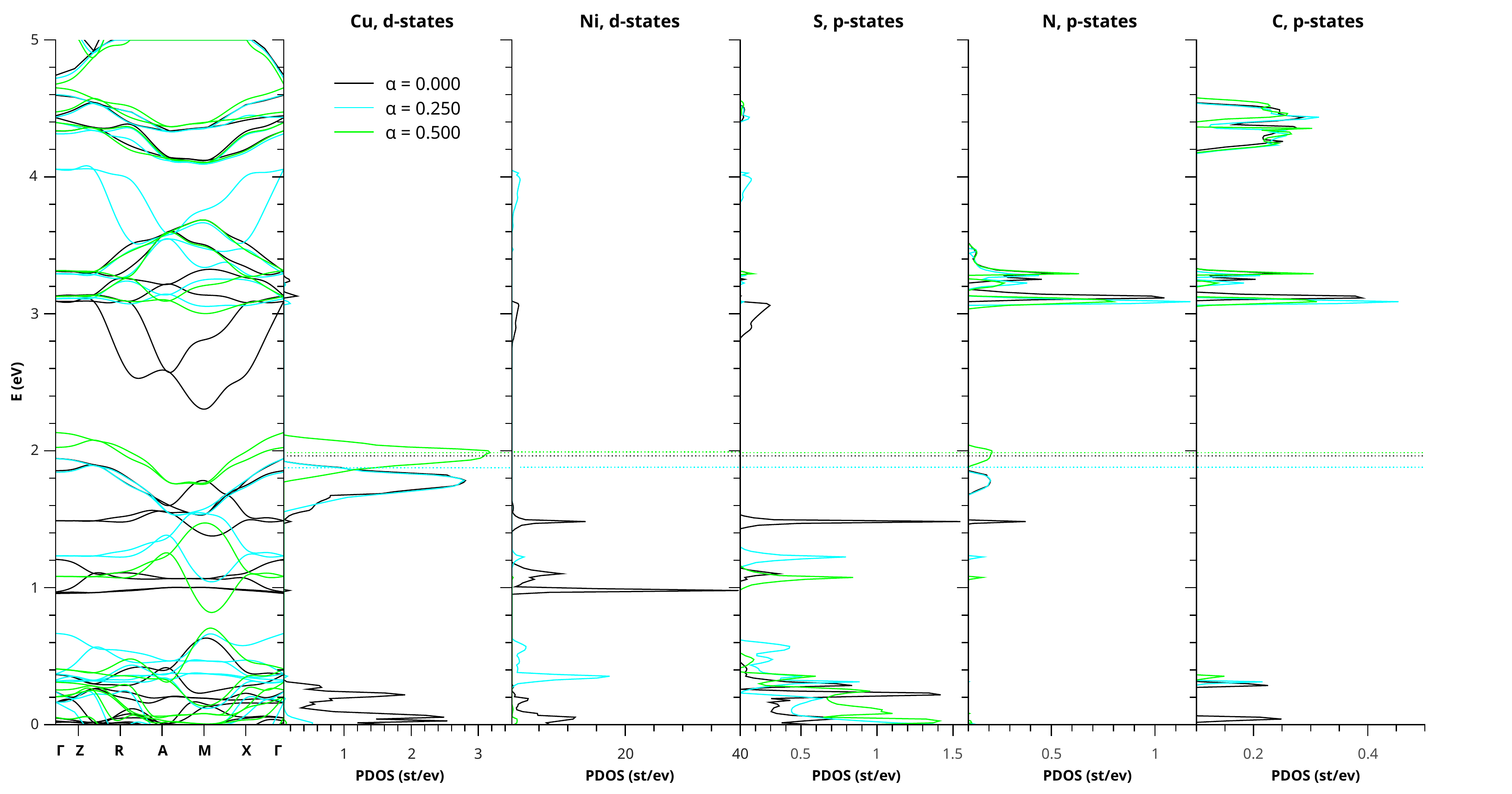}
\centering
\caption{(Color online) The electronic energy bands and PDOS in Cu[Cu(pdt)$ _{2} $] obtained with the hybrid functional PBE0, using several mixing parameters $ \alpha $. Dotted lines represent Fermi level.}
\label{fig:cu_ni_gs} 
\end{figure}

A significant difference in ground state properties is observed in Cu[Ni(pdt)$ _{2} $], compared with the material Cu[Cu(pdt)$ _{2} $]. 
Similarly to Cu[Cu(pdt)$ _{2} $], the top of the valence band is mainly represented by  d-states of Cu, but there are no p-states of S (figure~\ref{fig:cu_ni_gs}). Instead of the last ones, p-states of N are observed. There is a slight dependence of $d$-states of Cu on the mixing parameter $ \alpha $.  The material Cu[Ni(pdt)$ _{2} $] is the $p$-type semiconductor. The Fermi level obtained within the GGA-PBE approach is located a little above a top of the valence band. However, the increase of $ \alpha $ causes the lowering of the Fermi level with respect to the valence band edge.

The 3$d$-states of Ni are located deeper in the valence band and their positions strongly depend on $ \alpha $  (figure~\ref{fig:cu_ni_gs}). The 3$d$-states of Ni are also present in the conduction band. Their energy levels have risen with an increase of $ \alpha $.

\begin{figure}[!t]
\includegraphics[width=10cm]{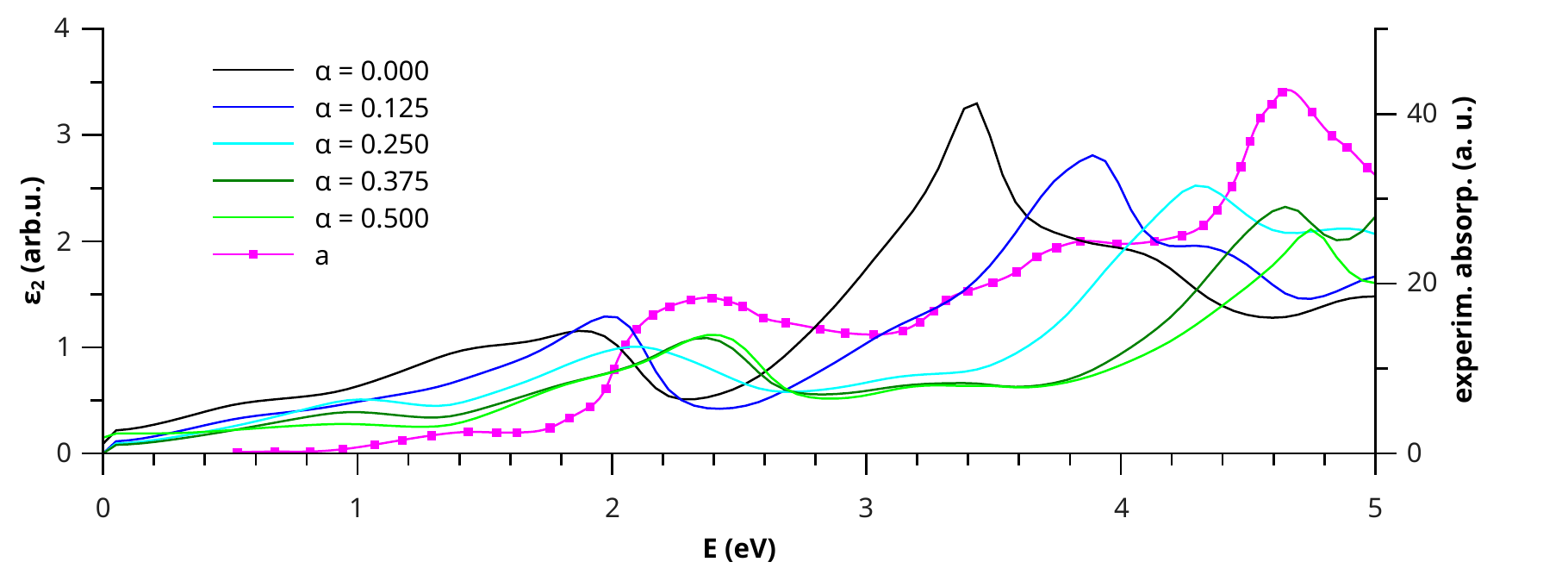}
\centering
\caption{(Color online) Imaginary part of the dielectric function $ \varepsilon_2^{\text{RPA}} $ for Cu[Ni(pdt)$ _{2} $], obtained  within the RPA on the base of hybrid functional PBE0, with several values of mixing parameter $ \alpha $,  in comparison with the experimental absorption spectrum,
	a -- \cite{doi:10.1021/ic802117q}
}
\label{fig:cu_ni_opt} 
\end{figure}

The imaginary parts of $ \varepsilon_2^{\text{RPA}} $, obtained with  values  $ \alpha $ = 0.25, 0.375 and 0.5, show a little peak located at the photon energy of about 1 eV (figure~\ref{fig:cu_ni_opt}). Comparing $ \varepsilon_2^{\text{RPA}} $ with the experimental absorption in the visible and ultraviolet (UV) part of spectrum, it can be noticed that the best agreement is achieved with the parameter value  $ \alpha $ = 0.50.

\subsection{Quasiparticle properties of Cu[Cu(pdt)$ _{2} $]}

The energy  scissor operator $ \Delta E_{g} $, needed for the BSE calculations, was derived as the difference of the band gaps obtained from the G0W0 and GGA approaches, which value is equal to 0.84 eV. 

\begin{figure}[!t]
\includegraphics[width=10cm]{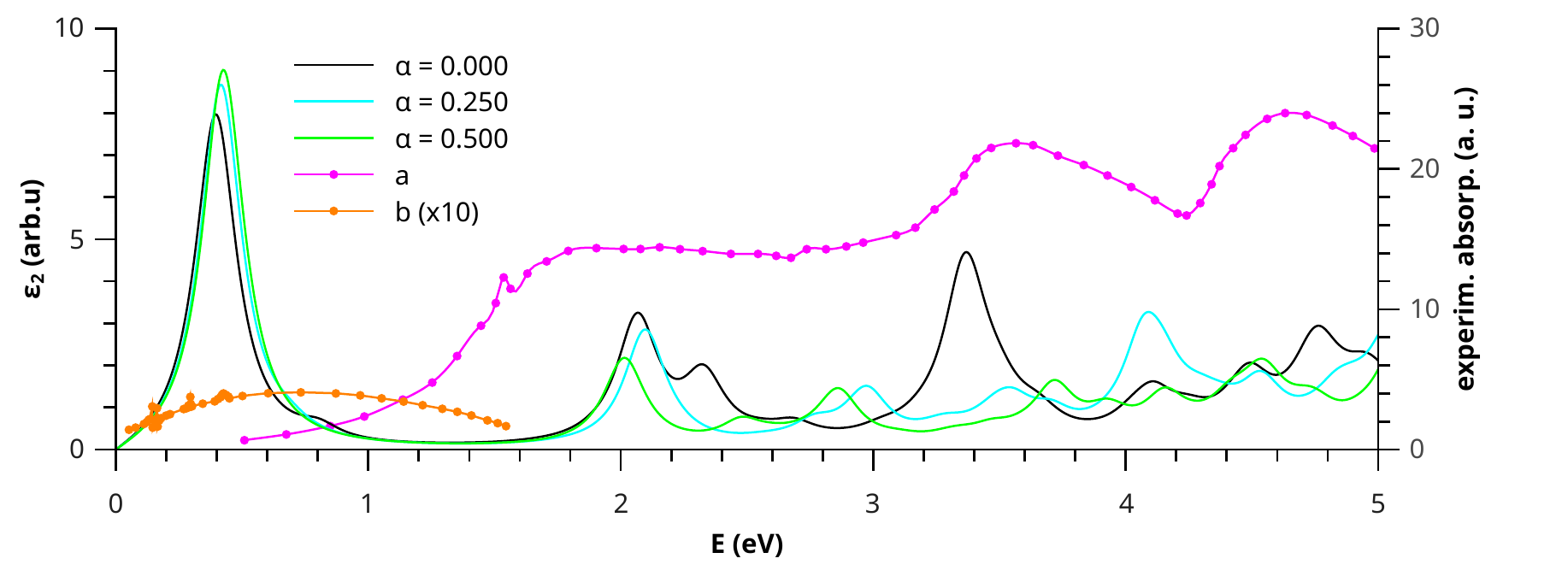}
\centering
\caption{(Color online) Imaginary part of the dielectric function $\varepsilon_2^{\text{BS}} $, obtained within the BSE approach, based on eigenstates, derived with the PBE0,  with several values for fraction $\alpha$ for Cu[Cu(pdt)$_{2}$], compared with the experimental absorption spectra,
a -- \cite{doi:10.1021/ic802117q},
b -- \cite{doi:10.1021/cm101238m}.
}
\label{fig:cu_cu_bse} 
\end{figure}

The behaviour of the imaginary part of the BSE dielectric function  $ \varepsilon_2^{\text{BSE}} $ (figure~\ref{fig:cu_cu_bse}) is very similar to $ \varepsilon_2^{\text{RPA}} $ (figure~\ref{fig:cu_cu_opt}). It has a strong peak at the photon energy of 0.5 eV, and its shape and energy are the same for all values of $ \alpha $. The first absorption peak in the visible spectrum, found within the BSE approach is  localized in the energy by 0.5 eV lower than the experimental one. A strong dependence of dielectric constant $ \varepsilon_2^{\text{BSE}} $ on the parameter $ \alpha $ is observed only in the UV photon energy region.

\subsection{Quasiparticle properties of Cu[Ni(pdt)$ _{2} $]}

The scissor energy $ \Delta E_{g} $, found within the G0W0 approach, which is built on the GGA-PBE ground-state eigenvalues and eigenfunctions, is equal to 1.10 eV.

\begin{figure}[!t]
\includegraphics[width=10cm]{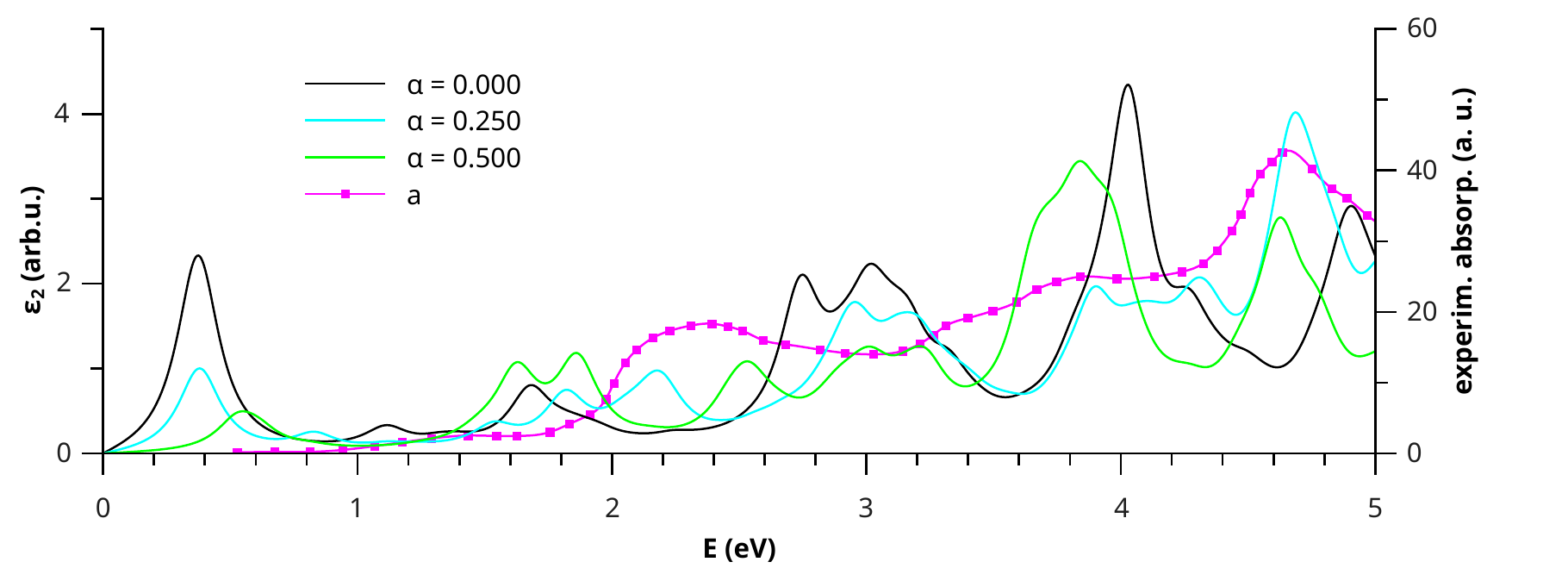}
\centering
\caption{(Color online) Imaginary part of the dielectric function $ \varepsilon_2^{\text{BSE}} $ obtained   based on eigenstates, derived from PBE0 approach, with several values of a fraction $ \alpha $, for Cu[Ni(pdt)$ _{2} $], compared with the experimental absorption spectrum,
a -- \cite{doi:10.1021/ic802117q}.
}
\label{fig:cu_ni_bse} 
\end{figure}

In case of GGA-PBE, $ \varepsilon_2^{\text{BSE}} $ has a strong peak located at 0.5 eV (figure~\ref{fig:cu_ni_bse}). It decreases rapidly and the peak energy shifts to the higher photon energy region with an increase of the parameter $ \alpha $. Similarly to the RPA results (figure~\ref{fig:cu_ni_opt}), the increase of the mixing coefficient $ \alpha $ has a significant impact in the visible and UV part of the absorption spectrum. In contrast to the results obtained for Cu[Cu(pdt)$ _{2} $] (figure~\ref{fig:cu_cu_bse}), in a visible energy region, the experimental and theoretical absorption spectra $ \varepsilon_2^{\text{BSE}} $ rise at higher photon energies (figure~\ref{fig:cu_ni_bse}). In the UV part, $ \varepsilon_2^{\text{BSE}} $ found with $ \alpha $ = 0.50 has two strong peaks, which agree well with the experiment.

\section{Discussion}

We have revealed that the use of the hybrid functional is of crucial importance in calculation of electronic and optical properties for materials containing the transition elements.
The importance of the removing SIE from exchange-correlation functional, for strongly correlated 3$d$ electrons, was established by a good agreement of ground state properties, obtained within PBE0 functional, with experimental data.
Moreover, the best agreement is achieved with the mixing coefficient $ \alpha $ which tends to 0.5. This value is two times higher than the default one, assumed in the PBE0 hybrid functional. Furthermore, we found that the removal of the SIE of the strongly correlated 3$d$ electrons of Cu is also essential. Recently, we became convinced that the SIE elimination  for Zn 3$d$ electrons improves the kinetic coefficients in the material ZnSe \cite{doi:10.1007/s11664-018-6068-1} which leads to better electronic energy bands in the ZnX (X: O, S, Se, Te) crystals, evaluated via combined scheme ``hybrid functional HSE06 + G0W0'' \cite{doi:10.21272/jnep.11(6).06018}. 
Recently, we performed calculations for materials that do not contain strongly correlated electrons. Therein the calculations that started from the level of the GGA  proved to be successful \cite{doi:10.5488/cmp.22.14701.2019}. However, for materials with strongly correlated 3$d$ electrons, we used  the hybrid PBE0 functional as a starting level.

The RPA and BSE approaches show the existence of absorption in the infrared (IR) energy region for both materials. However, in case of Cu[Cu(pdt)$ _{2} $], the energy and the shape of the peak in IR photon energies does not depend on the mixing parameter $ \alpha $. This is due to a simple shifting of band energy curves caused by the change of the mixing parameter $ \alpha $. In this material, a strong dependence of optical absorption is observed in the visible and UV photon energy region. This can also  be explained by shifting the bands near the Fermi level, while conduction bands remain frozen.
In Cu[Ni(pdt)$ _{2} $], material the conduction band energies are more sensitive to the fraction $ \alpha $ of exact Hartree-Fock energy in the hybrid exchange-correlation functional.
Though the 3$d$ states of Cu and Ni are located in the different intervals of energy, the dielectric functions $ \varepsilon_2^{\text{RPA}} $  of the Cu[Cu(pdt) and Cu[Ni(pdt) in IR photon energy part differ only slightly. 

The Fermi level in Cu[Cu(pdt)$ _{2} $] lies in the valence band, and this material behaves as a $p$-type semiconductor. The RPA and BSE approaches show similar results for dielectric function in IR, near IR and in the visible part of spectrum. Material Cu[Ni(pdt)$ _{2} $] is also a $p$-type semiconductor, but the Fermi level is located above the valence band in case of PBE and slowly sinks into the valence band with an increasing value of the mixing parameter. However, the difference between the dielectric constants, derived from the RPA and BSE approaches, is not significant.

The results of this work are obtained by combining different theoretical approaches. The calculated dielectric functions for the materials under study show a qualitative agreement with the experiment only in  certain intervals of the photon energy. It is clear that the results obtained here are dependent on the values of the mixing parameter $ \alpha $ in the hybrid PBE0 functional. The  scissor energy shift of the eigenenergies in the conduction band can also cause a calculation error for the dielectric function.

\section{Conclusions}

The electronic structure and optical properties of the porous coordination polymers Cu[Cu(pdt)$ _{2} $] and Cu[Ni(pdt)$ _{2} $] have been studied  based on different theoretical approaches. The strongly correlated 3$d$ electrons of Cu and Ni were treated by means of hybrid exchange-correlation functional PBE0 that partly removes the self-interaction error, which is very significant in case of 3$d$ electrons. For material Cu[Cu(pdt)]$ _{2}$ we found that the energies of the first peaks of the dielectric constants $ \varepsilon_2 $, found in the RPA and BSE approaches, are quite close. These peaks are well compared with the experimental optical absorption spectrum. The comparison of dielectric constants $ \varepsilon_2 $, evaluated with different values of the mixing factor $\alpha$, shows that the hybrid functional with a mixing coefficient $ \alpha  = 0.25$ provides a qualitative agreement with the measured data. The optical constants, found in PBE0 and BSE approaches, without incorporating Hartree-Fock energy into the exchange-correlation energy functional ($\alpha  = 0$), show the largest deviation from the measured values.

\newpage
\ukrainianpart

\title{Дослідження пористих координаційних полімерів Cu[Cu(pdt)$_{2}$] та Cu[Ni(pdt)$_{2}$] за допомогою гібридного функціоналу}
\author{С.В. Сиротюк, Ю.В. Клиско}
\address{Національний університет ``Львівська політехніка'',  вул. Степана Бандери, 12,\\ 79013,  Львів, Україна}

\makeukrtitle

\begin{abstract}
\tolerance=3000%
Проведено ab initio дослідження двох пористих координаційних полімерів Cu[Cu(pdt)$_{2}$] і Cu[Ni(pdt)$_{2}$]. Закони дисперсії та парціальні густини станів були отримані з використанням гібридного функціоналу PBE0. Знайдені результати виявили, що розглянуті матеріали є виродженими напівпровідниками p-типу. Проаналізовано значний вплив сильних кореляцій 3$d$-електронів нікелю та міді на електронні властивості матеріалів. 
Встановлено, що для матеріалу Cu[Cu(pdt)$ _ {2} $] також важливе врахування сильних кореляцій $d$-електронів, які ми врахували за допомогою гібридного функціонала обмінно-кореляційної енергії PBE0. 
Квазічастинкові характеристики були отримані за методом функції Гріна (GW) та на основі рівняння Бете-Солпітера. Останній був використаний для дослідження екситонних властивостей вироджених напівпровідників. Уявна частина діелектричної функції була отримана в межах наближення хаотичної фази, а також з явним урахуванням взаємодії електрона і дірки, імплементованої у рівнянні Бете-Солпітера.
\keywords полімери, пористі матеріали, металоорганічні сполуки, гібридний функціонал, метод~GW, рівняння Бете-Солпітера

\end{abstract}

\end{document}